\RequirePackage{fix-cm}
\documentclass[smallextended]{svjour3}       % onecolumn (second format)
\smartqed  % flush right qed marks, e.g. at end of proof
\usepackage{graphicx}
\usepackage{amsmath}
\usepackage{amsfonts}
\usepackage{amssymb}

\usepackage{color}
\usepackage{bbm}
\usepackage{mathtools}
\usepackage{empheq}
\usepackage{physics}
\usepackage[active]{srcltx}
\usepackage{setspace}
\usepackage{graphicx}% Include figure files
\usepackage{dcolumn}% Align table columns on decimal point
\usepackage{bm}% bold math
\usepackage{epsfig}
\usepackage{epstopdf}
\usepackage{esint}
\usepackage{caption}
\usepackage{subcaption}
\usepackage[backend=biber,style=numeric,sorting=none]{biblatex}

\usepackage{siunitx}

\begin{document}

\title{Ergontropic Dynamics%\thanks{Grants or other notes
}
\subtitle{Contribution for an Extended Particle Dynamics}


\author{Mario J. Pinheiro         \and
%        Second Author %etc.
}


\institute{Mario J. Pinheiro \at
              Department of Physics, Instituto Superior T\'{e}cnico-IST, Universidade de Lisboa, Av. Rovisco Pais, 1049-001 Lisboa Codex, Portugal \\
              Tel.: +351-1-21-841-9322\\
              Fax:  +351-1-21-846-4455\\
              \email{mpinheiro@tecnico.ulisboa.pt}           %  \\
           \and
}

\date{Received: date / Accepted: date}

\maketitle

\begin{abstract}
A vast concourse of events and phenomena occur in nature that may be interrelated by a entropy-maximization technique that provides a comprehensible explanation of a range of physical problems, integrating in a new framework the universal tendency of energy to a minimum and entropy to a maximum. The outcome is a modification of Newton's dynamical equation of motion, grounding the principles of mechanics on the concepts of energy and entropy, instead on the usual definition of force, integrating into a consistent framework the description of translation and vortical motion. The new method offers a fresh approach to traditional problems and can be applied with advantage in the solution of variational problems.

\end{abstract}

\section{Introduction}
\label{intro}
The classical mechanistic description of Sir Isaac Newton and Pierre-Simon Laplace presupposes the reversibility of time, i.e., time follows it course (irreversibly) but it didn't affect the possibility of the reverse motion. Historically, the mechanistic theory was called into question when the first anomalies appeared in heat experiments, namely, the fact that an warm body transmit heat to a cold body until their temperature equalize, but never the reverse happened, indicating the first striking example of irreversible phenomena. However, in Newtonian, or Galilean mechanics, an observer $O$ on an inertial reference frame $R$, recognizes an equilibrium state $A$ of a mechanical system $M$ (that we may visualize as an ideal fluid or a ``rigid body"), assigning to this state $A$ an energy $E$ and a number $N$ of particles constituents (by virtue its mass). And, from these assumptions, it follows the thermodynamic implications of the isotropy and homogeneity of Newtonian space-time.

Here, we present an approach differing in fundamental aspects from the standard treatments, particularly not deriving irreversible thermodynamics from the reversible microscopic dynamics. By the contrary, we relate dynamics and thermodynamics based on the concept of energy, work, and entropy. At the core of the present theory is the second law of thermodynamics which has a form of causality as a universal property. Working out the general method here proposed, at the end, a system can be described in terms of dynamical quantities and free energy.

In a previous publication~\cite{Pinheiro_2013} we have shown that rotational motion and free energy gradients imply dissipation of energy, are sources of irreversible phenomena. We may place the gradient of rotational energy in the class of transport of angular momentum problems. 

In this proposed program, we develop a standard technique for treating a physical system on the basis of an information-theoretic framework previously developed~\cite{Landau,Jaynes,Pinheiro_02,Pinheiro_04,Pinheiro_2013}. The proposed technique starts with the total entropy of the system composed of $N$ particles (or bodies), then it is applied the method of the Lagrange multipliers for the differential of entropy (0-form) $d \overline{S}=\nabla_k \overline{S} dx^k$ $(k=x,y,z)$; finally, the total entropy is inserted on the fundamental equation of entropy written under differential form, $\nabla_k U= T \nabla_k \overline{S} - F_k$, summing over the all ensemble of $N$ particles.
This method lead to a set of two first-order differential equations revealing the same formal symplectic structure shared by classical mechanics and thermodynamics~\cite{Peterson_1979,TMP_Sergeev_2008}. When the maximization of entropy is seek, it results on the well-known equations of (electro)dynamics (if electromagnetic entities enter into the system). This work leads to a reformulation of mechanics and electrodynamics, in particular, giving a new definition of the equation of motion that explicitly displays irreversibility. 

Consider here a simple dynamical system constituted by a set of $N$ discrete interacting mass-points (although macroscopic) 
$m^{(\alpha)}$ ($\alpha=1,2,...,N$) with $x_i^{\alpha}$ and
$v_i^{\alpha})$ ($i=1,2,3; \alpha=1,...,N$) denoting the
coordinates and velocities of the mass point in a given inertial
frame of reference. The inferior Latin index refers to the
Cartesian components and the superior Greek index distinguishes
the different mass-points.

Following the line of thought proposed in Ref.~\cite{Pinheiro_04} the total energy of a, let's say, mechano-electric system constituted by $N$ interacting particles, which includes here, for the purpose of illustration, the gravitational potential $\phi^{(\alpha)}$ for the species $\alpha$, and the gravitational interaction $\phi^{(\alpha,\beta)}$ between specie $\alpha$ and $\beta$ (all the other terms are well-known in the technical literature) is given by:

\begin{eqnarray}\label{eq1}
E=&&\sum_{\alpha=1}^N [ U^{(\alpha)}+ \frac{p^{(\alpha)2}}{2 m^{(\alpha)}} +\frac{J^{(\alpha)2}}{2I^{(\alpha)}} + q^{(\alpha)} V^{(\alpha)} \nonumber\\
&& - q^{(\alpha)} (\mathbf{A}^{(\alpha)} \cdot \mathbf{v}^{(\alpha)}) +m^{(\alpha)} \phi^{(\alpha)} (\mathbf{r}) 
+ m^{(\alpha)}\sum_{\beta=1}^N \phi^{(\alpha,\beta)} ].
%\end{multlined}
\end{eqnarray}

The next fundamental task is the construction of the invariant entropy function $\overline{S}$, as a function of the, so-far, {\it known} fundamental invariants in a closed system. Hence, $\overline{S}=\overline{S}(E,\pmb{p},\pmb{L})$, and we must construct the entropy as a function of the linearly independent components via the Lagrange multiplier form of entropy (LMFE) given by

\begin{multline}\label{eq2}
\overline{S} = \sum_{\alpha=1}^N  
[S^{(\alpha)} ( E^{(\alpha)}
- \frac{(\mathbf{p}^{(\alpha)})^2}{2 m^{(\alpha)}}
- \frac{(\mathbf{J}^{(\alpha)})^2}{2 I^{(\alpha)}}
- q^{(\alpha)} V^{(\alpha)}
+ q^{(\alpha)} (\mathbf{A}^{(\alpha)} \cdot \mathbf{v}^{(\alpha)}) \\- U^{(\alpha)}_{mec} )
+(\mathbf{a} \cdot \mathbf{p}^{(\alpha)}
+\mathbf{b} \cdot ([\mathbf{r}^{(\alpha)}
\times \mathbf{p}^{(\alpha)}] + \mathbf{J}^{(\alpha)}) ]
\end{multline}
or
\begin{equation}
\overline{S}= \sum_{\alpha=1}^N \mathfrak{S}^{(\alpha)}.    
\end{equation}

Here, $\mathbf{a}$, and $\mathbf{b}$, are Lagrange multipliers vectors, related to the velocity of translation $\mathbf{v}_e=T \mathbf{a}$, and the velocity of rotation $\pmb{\omega}=T\mathbf{b}$. The method of Lagrange multipliers is applied to find extremal values of a function of several variables subject to one, or more, constraints. 
The absolute temperature is denoted by $T$, more realistically interpreted as the kinetic temperature, the equipartition expression in momenta. Thus, for an isolated extended particle in thermodynamic equilibrium, the translational and rotational motion is only possible as a rigid body~\cite{Landau}. 

The entropy, assumed an invariant to the observer $O$ in a given reference frame (possibly under rotations), is an invariant function, with 8 linearly independent components: mass $M$, energy $E$, total linear momentum $\pmb{p}$, total angular momentum $\pmb{L}$.

The introduction of the electrostatic and magnetic energy term were also used in a previous publication~\cite{Pinheiro_04}, were some consequences of this formulation were explored in electrodynamics. The additional terms included in the LMFE are the total linear momentum and the total angular momentum about the origin, both with respect to a definite initial reference frame:
\begin{subequations} \label{eq1a}
\begin{align}
\sum_{\alpha} \mathbf{p}_{\alpha} & = \mathbf{P} = const, \\ 
\sum_{\alpha} ([\mathbf{r}_{\alpha} \times \mathbf{v}_{\alpha}]+ \mathbf{J}_{\alpha}) & = \mathbf{J} = const,
\end{align}
\end{subequations}
with $\mathbf{J}$ denoting the total sum of the orbital and spin angular momentum, both equations due to the principles of energy and angular momentum conservation in an isolated body, constituted by all $\alpha=1,2,...N$ particles, or bodies, all at the same temperature to avoid, at this level of description, mathematical complexity. Eqs.~\ref{eq1a}(a,b) are the constraints equations referred above. The aim of the method of the Lagrange multipliers is to allow the determination of a particular set of points where a given function (here, the function $\overline{S}$) have minimal or maximal values.

Extremum conditions imposed on entropy or internal energy not only
provide criteria for the evolution of the system but determine as
well the stability of thermodynamic systems at equilibrium. It has been shown~\cite{Lavende:74} that a state of mechanical
equilibrium (or in the absence of convective motion) is onset when the entropy increase with distance from equilibrium, such as:
\begin{equation}\label{eq2a}
\frac{\partial \overline{S}}{\partial
\mathbf{r}} > 0.
\end{equation}
Applying this concept to Eq.~\ref{eq2}, it is obtained the entropic flux in space, a kind of generalized force $X_{\alpha}$
~\cite{Prigogine:71}, with its value given by
\begin{equation}\label{eq3}
\frac{T}{2} \frac{\partial \overline{S}}{\partial \mathbf{r}_i^{(\alpha)}} =
\nabla_{r_i} U^{(\alpha)} + m_i \frac{d v_i^{(\alpha)}}{dt} + \frac{1}{2}
\nabla_{r_i} (\pmb{\omega} \cdot \mathbf{J}_i),
\end{equation}
where 
\begin{equation}
U^{(\alpha)}_{eq} = m^{(\alpha)} \phi^{(\alpha)} (\mathbf{r}) + m^{(\alpha)}\sum_{\beta=1}^N \phi^{(\alpha,\beta)},
\end{equation}
represents the internal energy of a given 
mass-point $\alpha$ at equilibrium. When seeking for an extremal value, imposing $\nabla \overline{S}=0$, it gives the fundamental hydrodynamic equation:

\begin{multline}\label{eq4}
m^{(\alpha)} \frac{d v_i^{(\alpha)}}{dt} = - \nabla_r
(m^{(\alpha)} \phi^{(\alpha)}) - m^{(\alpha)} \nabla
\sum_{\beta=1}^N \phi^{(\alpha,\beta)}- \frac{1}{2} \pmb{\nabla}_{r^{(\alpha)}} (\pmb{\omega} \cdot \mathbf{J}^{(\alpha)}).
\end{multline}

In conformity with the definition of temperature, we look at the gradient of the total entropy in
momentum space multiplied by factor $T$, which gives
\begin{equation}\label{6}
T \frac{\partial \overline{S}}{\partial \mathbf{p}_i} = \{
-\frac{\mathbf{p}_i}{m_i} + \frac{q_i}{m_i}\mathbf{A} +
\mathbf{v}_{rot} + [\pmb{\omega} \times \mathbf{r}_i] \}.
\end{equation}

The maximization of Eq.~\ref{6} leads to the total (canonical)
momentum, given by
\begin{equation}\label{eq7}
\mathbf{p}_i = m_i \mathbf{v}_{rot} + m_i [\pmb{\omega} \times
\mathbf{r}_i] + q_i \mathbf{A}.
\end{equation}
However, in non-equilibrium conditions, the exact canonical momentum for each particle is
\begin{equation}\label{eq7a}
\mathbf{p}=m\mathbf{v}_{rot}+m[\pmb{\omega} \times \mathbf{r}]+q\mathbf{A}-mT \pmb{\nabla_p} \overline{S}.
\end{equation}

\begin{figure}
  \includegraphics[width=4.250 in]{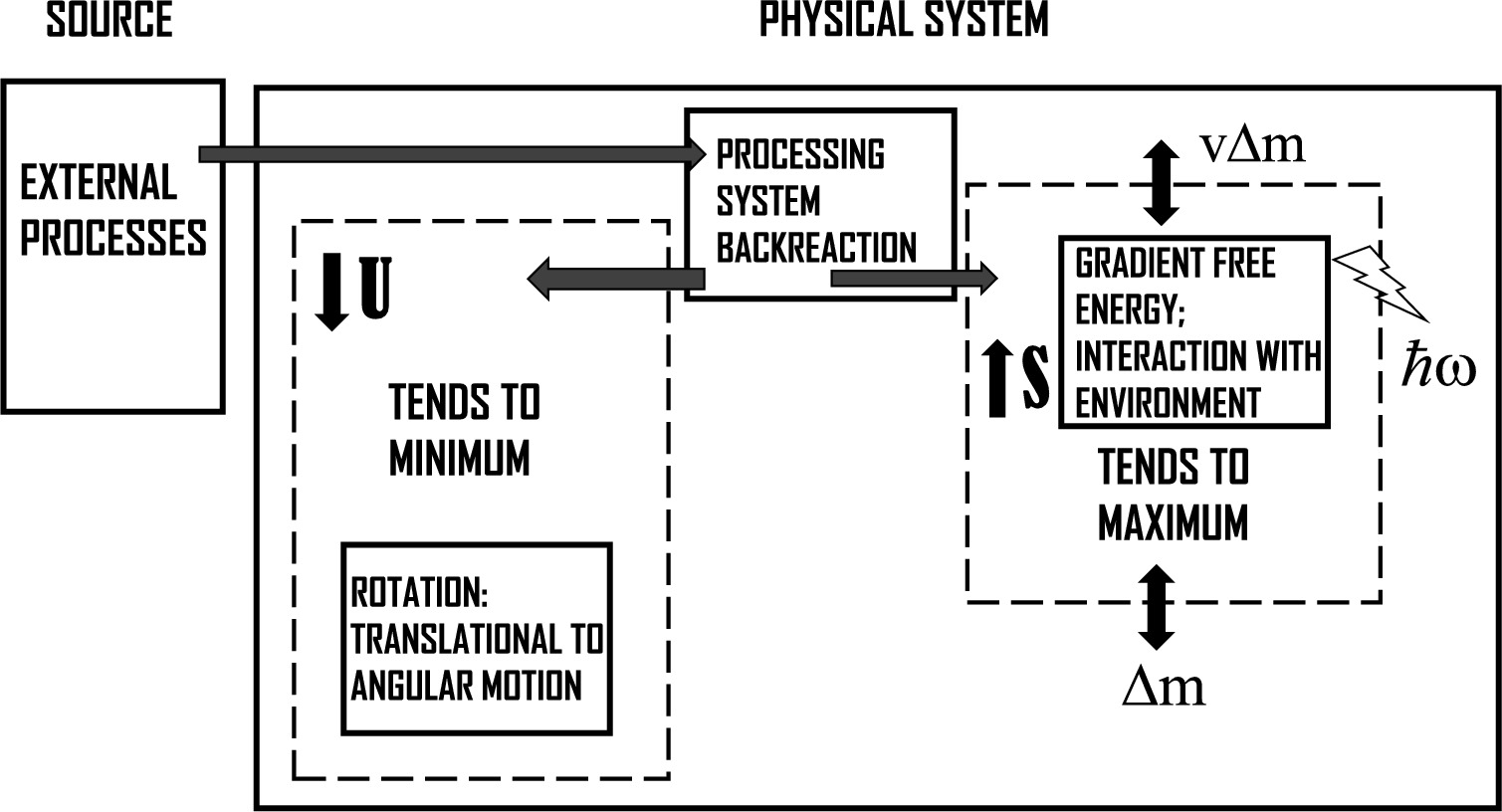}\\
  \caption{System process chain – a physical system has, at least, two ways to process the external action (it is not enslaved to it), by means of two processes chain: energy tends to a minimum by means of conversion of linear or transversal motion to rotational motion; Entropy tends to a maximum by means of decreasing free energy (e.g., ejection of mass, acoustic or electromagnetic radiation).}\label{fig1a}
\end{figure}

This variational approach, based on the entropy-energy principle, points to the dominance and interplay of the fundamental dualism:
\begin{description}
\item[Entropy]: i) do not explicitly depend on energy; ii) is dependent of the particles configuration (or collectively dependent); iii) it has a characteristic speed of "entropic propagation" (equal at maximum to the speed of light);
\item[Energy]: i) decay over time and space; ii) propagates with characteristic speed, at maximum the speed of light.

\end{description}
In the course of advancing knowledge the two have exchanged places on the understanding of natural processes. Entropy dictates how the all process evolves, while energy merely does the bookkeeping. Fig.~\ref{fig1a} shows the link between propensities and variables.

It was shown that in the case of an extended rotating charged sphere the canonical momentum is given by
\begin{equation}\label{eq7b}
\mathbf{p}=m\mathbf{v}+q\mathbf{A}+\frac{1}{2}\gamma I \nabla^2 \mathbf{A}.
\end{equation}
Here, $\gamma$ is the gyromagnetic ration and $I$ is the moment of inertia~\cite{Young_1975}. We may associated to the term $\nabla^2 \mathbf{A}$ a spread field in space such as $-\frac{\mathbf{A}}{\lambda_e}$, with $\lambda_e$ denoting the Compton wavelength, in which case it would appear a correction to the electron observed charge $e_R$. This correction results from the collective action of all the other charged electrons on the test particle and the correction, in a crude estimation, would be of the order, $e_r=e(1-\alpha^2/16 \pi^2)$. In view of this, we may observe that it is not correct to adopt the usual definition of a particle as a system without extension, as a system of negligible size. Continuous mechanics is a far-more general theory than the usual mass-point mechanics~\cite{Truesdell_1952}.

Now, let us assume that the total energy $E$ is constant. We then obtain (letting down the index $\alpha$) by taking the gradient of the Eq.\ref{eq1}:
\begin{equation}
\pmb{\nabla}_r U=m \mathbf{v} \cdot \pmb{\nabla} \mathbf{v}-\pmb{\nabla} \left( \frac{\mathbf{J}}{2I}-\pmb{\omega} \cdot \mathbf{J} \right)-q \pmb{\nabla} V+q\pmb{\nabla}(\mathbf{A} \cdot \mathbf{v}).
\end{equation}

You'll notice that the non-translational energy of the system $U$ (or the internal energy) replaces the the total energy, determines the actual existence of mechanical states.

Applying the present formalism in a previous work~\cite{Pinheiro_04}, we obtained the ponderomotive forces acting on a charged particle. For a neutral particle (or rigid body) we can obtain from Eq.~\ref{eq3}, when there is no flux of entropy, a kind of extended fundamental equation of dynamics for a given species $(\alpha)$:
\begin{equation}\label{eq3a}
m^{(\alpha)} \frac{d \mathbf{v}^{(\alpha)}}{dt} = \mathbf{f}^{(\alpha)} - \frac{1}{2} \pmb{\nabla}_{r^{(\alpha)}} (\pmb{\omega} \cdot \mathbf{J}^{(\alpha)}).
\end{equation}
Eq.~\ref{eq3} has a new term because the body possess an intrinsic angular momentum, and this expression can be integrated according to the theorem work-energy in order to obtain the mechanical energy of the system, $E_{mec}=K+U+I\omega^2/2$. There is an interesting aspect related to Eq.~\ref{eq3}: classical problems of mechanics, as for example, rigid body rolling down an inclined plane (see, e.g., p. 97 of Ref.~\cite{Lamb}) may be solved in an integrated manner.
At this stage, it is worthwhile to refer that our procedure includes the effect of angular momentum (through Eq.~\ref{eq2}), as it should be in a consistent theory, according to Curtiss~\cite{Curtiss 1956}.

\begin{figure}
  \includegraphics[width=3.0 in]{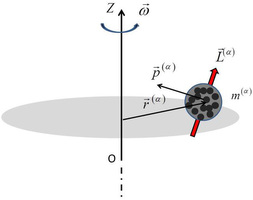}\\
  \caption{An assortment of particles of mass $m^{(\alpha)}$, in rotational motion around an axis OZ with angular velocity $\pmb{\omega}$, where $\mathbf{r}^{(\alpha)}$ denotes the position vector
relative to a fixed reference frame, $\mathcal{R}$,
$\mathbf{p}^{(\alpha)}$ denotes the total momentum (particle + field) and
$\mathbf{L}^{(\alpha)}$ denotes the total angular momentum of the
particle. Courtesy from Nature research Journal, Ref.1.}\label{fig2}
\end{figure}

\subsection{Forces in an accelerated frame}

Now remark that the term $\nabla_r (\pmb{\omega} \cdot \mathbf{J})$ may be developed in the following manner (from now on we don't refer to the set of particles and discard the supra-index $(\alpha)$):

\begin{multline}\label{eq8}
\frac{1}{2} \pmb{\nabla_r} (\pmb{\omega} \cdot \mathbf{J}) = \frac{1}{2}[\pmb{\omega} \times \mathrm{curl} \; \mathbf{J}] + \frac{1}{2}(\pmb{\omega} \cdot \pmb{\nabla}) \mathbf{J}+\frac{1}{2}[\mathbf{J} \times \mathrm{curl}\; \pmb{\omega}] +\frac{1}{2}(\mathbf{J} \cdot \pmb{\nabla}) \pmb{\omega}.
\end{multline}

Within this new formulation, the transformation from an inertial coordinate system to a rotating system shows that the so called ``fictitious forces", although certainly not related to interacting fields, they are the outcome of the transport of angular momentum, they are real forces in the rotating medium. We should have in mind that there is active transformations (implying its motion, strongly or weakly coupled to other systems) and passive transformations (method of describing transitions to different reference systems, e.g., Galileo or Lorentz transformations). Any mathematical procedure can be reliable unless it is completely understood the purpose and how it should be applied, otherwise becomes a source of confusion.

In order to justify what we have state above, one therefore has to work out the resultant term $\nabla_r (\pmb{\omega} \cdot \mathbf{J})$ in Eq.~\ref{eq3a}, developing it in the following manner:
\begin{equation}\label{eq8}
\frac{1}{2} \pmb{\nabla_r} (\pmb{\omega} \cdot \mathbf{J}) = \frac{1}{2}[\pmb{\omega} \cp \curl{\vb{J}}] + \frac{1}{2}(\pmb{\omega} \cdot \pmb{\nabla}) \mathbf{J}+
\frac{1}{2}[\mathbf{J} \cp \curl{\pmb{\omega}}] +\frac{1}{2}(\mathbf{J} \cdot \pmb{\nabla}) \pmb{\omega}.
\end{equation}
We may test the above Eq.~\ref{eq8} with the well-known problem of rotating coordinate frame, without translation, considering a rotating system $(x',y',z')$ whose origin coincides with the origin of an inertial system $(x,y,z)$. We suppose in addition, that the $z$ and $z'$ axes always coincide and, therefore, the angular velocity of the rotating system, $\pmb{\omega}$, lies along the $z$ axis. We take $\mathbf{J}=I\pmb{\Omega}$ and $\pmb{\Omega}=\pmb{\omega}$, since we consider the ``body" in rotational motion around the $z$ axis. An easy calculation (considering $I_z=mr^2$) show that
\begin{equation}\label{eq9}
\curl{\mathbf{J}} = 2 m\mathbf{v}_{rot},
\end{equation}
where $\mathbf{v}_{rot}$ denotes the particle velocity in the rotating frame. Also, following the analogy
\begin{equation}\label{eq10}
\curl{\mathbf{v}} = \mathbf{\Omega}=2\pmb{\omega},
\end{equation}
and using a mathematical identity, we obtain
\begin{equation}\label{eq11}
\curl{} \left( \frac{\mathbf{J}}{r^2} \right)=\frac{1}{r^2} \curl{\vb{J}}+ \left[\grad{} \left( \frac{1}{r^2} \right) \times \mathbf{J} \right],
\end{equation}
which gives
\begin{equation}\label{eq10a}
\curl{} \left( \frac{\mathbf{J}}{r^2} \right)=\frac{2m}{r^2}(\mathbf{v}_{rot})+[\pmb{\omega} \times \mathbf{r}].
\end{equation}
Then
\begin{equation}\label{eq10b}
\frac{1}{2} \left[\mathbf{J} \times \curl{} \left( \frac{\mathbf{J}}{r^2} \right)\right]=m \pmb{\omega} \times [\mathbf{v}_{rot} + [\pmb{\omega} \times \mathbf{r}]],
\end{equation}
and, therefore, adding together all the terms, it is finally obtained the well-known formula of Newtonian dynamics:
\begin{equation}\label{eq17}
\left( m\frac{d\mathbf{v}}{dt} \right)_{rot}=\mathbf{F}^{ext}-2m[\pmb{\omega} \times \mathbf{v}_{rot}]-m(\pmb{\omega} \times [\pmb{\omega} \times \mathbf{r}]).
\end{equation}
We stress that in Eq.~\ref{eq17} the force term is calculated in the non-inertial frame inserting the acceleration force inside the symbol $(...)_{rot}$.

The Euler's force term ($d\pmb{\omega}/dt \times \mathbf{r}$) does not appear in Eq.~\ref{eq17} because the assumption behind Eqs.~\ref{eq3}-~\ref{eq7} is of extremum of entropy and $d \omega/dt=0$. It is worth to note that Eq.~\ref{eq3a} gives the force measured on the moving reference frame for a given part of mass $m^{(\alpha)}$ pertaining to an extended mass. In addition, it shows that the force exerted on an extended body maybe quite different from the one that it is usually expected, based on inferences that follow from the point particle dynamics. In particular, the changing of inertial moment of the extended particle along its motion, may lead to new effects.

This means that Eq~\ref{eq3a} is more general than the well-known Eq.~\ref{eq17}, valid for an ideal particle, provided that the constriction Eq.~\ref{eq9} is obeyed for an extended body.

\section{The Ergontropic Equation of Dynamics}

Taking the gradient of $\overline{S}$ in Eq.~\ref{eq2}, we obtain
\begin{equation}\label{eq18}
\grad{\overline{S}}_{r_i}=\pdv{S_i}{U_i}\pdv{U_i}{\mathbf{r}_i}+m\mathbf{a}_i \vdot \pdv{\mathbf{v}_i}{\mathbf{r}_i}+\mathbf{b} \vdot \pdv{\mathbf{r}_i}([\mathbf{r}_i \cp \mathbf{p}_i]+\mathbf{J}_i).
\end{equation}

Here, by definition of temperature, $\frac{1}{T}=\pdv{S_i}{U_i}$, and $\mathbf{v}_e=\mathbf{a} T$, $\mathbf{\pmb{\omega}}=\mathbf{b}T$. 

Then (now dropping the i-index to simplify), it is obtained
\begin{equation}\label{eq19}
T \grad{\overline{S}}_{r}=\pdv{U}{\mathbf{r}}+m\mathbf{v}_e \vdot \pdv{\mathbf{v}}{\mathbf{r}}+\pdv{\mathbf{r}}(\pmb{\omega} \vdot [\mathbf{r} \cp \mathbf{p}])+ \pmb{\omega} \vdot \pdv{\mathbf{J}}{\mathbf{r}}.
\end{equation}
The internal energy is the summing up of the kinetic energy with the potential energy (e.g., gravitational, electromagnetic) and, therefore,
\begin{equation}\label{eq20}
\pdv{U}{\mathbf{r}}=-m\mathbf{v} \vdot \pdv{\mathbf{v}}{\mathbf{r}}-m\pdv{\Phi}{\mathbf{r}}.
\end{equation}
Hence,
\begin{equation}\label{eq21}
T \grad{\overline{S}}_{r} = -m\mathbf{v} \vdot \pdv{\mathbf{v}}{\mathbf{r}} - m\pdv{\Phi}{\mathbf{r}} + m\pdv{\mathbf{r}}(\mathbf{v} \vdot (\mathbf{v}_e + [\pmb{\omega} \cp \mathbf{r}])) + \pdv{\mathbf{r}}(\pmb{\omega} \vdot \mathbf{J}).
\end{equation}
or,
\begin{equation}\label{eq22}
T \grad{\overline{S}}_{r} = -m \mathbf{v} \vdot \pdv{\mathbf{v}}{\mathbf{r}} - m\pdv{\Phi}{\mathbf{r}} + m\pdv{\mathbf{r}}(\mathbf{v} \vdot \mathbf{v}) + \pdv{\mathbf{r}}(\pmb{\omega} \vdot \mathbf{J}).
\end{equation}
Finally, getting them together, we get
\begin{equation}\label{eq23}
T \grad{\overline{S}}_{r} = m\mathbf{v} \vdot \pdv{\mathbf{v}}{\mathbf{r}} - m\pdv{\Phi}{\mathbf{r}} - \pdv{\mathbf{r}} \left( \frac{J^2}{2I} - \pmb{\omega} \vdot \mathbf{J} \right).
\end{equation}
The first term on the rhs of Eq.~\ref{eq23} is the convective term and we know that
\begin{equation}\label{eq24}
\frac{d\mathbf{v}}{dt} = \pdv{\mathbf{v}}{t} + \mathbf{v} \vdot \pdv{\mathbf{r}}{\mathbf{v}}.
\end{equation}
However, the net force acting on the volume element of the body causes the acceleration $\pdv{\mathbf{v}}{t}$, in which case we should replace
\begin{equation*}
  -m\mathbf{v} \vdot \pdv{\mathbf{v}}{\mathbf{r}} \to m\pdv{\mathbf{v}}{t}.
\end{equation*}
This amount to the recognition of the principle of equivalence, and for an observer sit on the extended particle center of mass, he is at rest. Then it results
\begin{equation}\label{eq25}
T \grad{\overline{S}}_{r} = - m\pdv{\mathbf{v}}{t} - m\pdv{\Phi}{\mathbf{r}} - \pdv{\mathbf{r}} \left( \frac{J^2}{2I} - \pmb{\omega} \vdot \mathbf{J} \right).
\end{equation}
We can rewrite Eq.~\ref{eq25} under the more suitable form
\begin{equation}\label{eq26}
 m\pdv{\mathbf{v}}{t} = - m\pdv{\Phi}{\mathbf{r}} - \pdv{\mathbf{r}} \left( \frac{J^2}{2I} - \pmb{\omega} \vdot \mathbf{J} + TS \right),
\end{equation}
assuming, to simplify, that the process has no thermal gradients. Finally, and for more practical uses, we introduce instead the free energy term, in which case the final form of the Newtonian dynamics become more consistently represented in the complete form:
\begin{equation}\label{eq27}
 \boxed{m\pdv{\mathbf{v}}{t} = - m\pdv{\Phi}{\mathbf{r}} - \pdv{\mathbf{r}} \left( \frac{J^2}{2I} - \pmb{\omega} \vdot \mathbf{J} - F \right).}
\end{equation}
This equation describes the combined motion of rotation and translation, with a rotation about an {\it instantaneous centre} with angular velocity $\pmb{\omega}(t)$. The free energy term inside Eq.~\ref{eq27} ensures the tendency of the free energy to the minimum, tending to the smallest possible value, connected to the term $-TS$ that actuates to give the maximum possible value, that is, ensuring that all possible states can be realized equally.

Such a formulation suggests that the external forces do not dictate the fate of a physical system {\it per se}, as in fact, due to entropy (and information entropy, via Shannon's theorem) related phenomena, the physical system can organize itself, can tend towards novel behavioural patterns. Here I also wish emphasize that the action-to-reaction law needs a reinterpretation, different from the deprecated form as appearing in standard text books. The reaction follows the action in a time lag that depends of the entropy (or information) gradients; the release from constraints follow the tendency towards new arrangement of the parts of the system.

\subsection{Rolling body on an inclined plan}

We may check Eq.~\ref{eq3} in a standard example of classical mechanics: a rigid body of mass $M$ rolling down an inclined plane making an angle $\theta$ with the horizontal (see, e.g., p. 97 of Ref.~\cite{Lamb}). Eq.~\ref{eq4} can be applied to solve the problem, with $\omega=0$ (there is no rotation of the frame of reference) and considering that only the gravitational force acts on the rolling body, with inertial moment relative to its own center of mass given by $I_c=\beta MR^2$. Hence, we obtain:
\begin{equation}\label{eq7c}
M \ddot{x}= Mg \sin \theta - \partial_x w.
\end{equation}
Here, $w \equiv \frac{(\mathbf{J_c})^2}{2 I_c}$. Assuming that the x-axis is directed along the inclined plane, and considering that the angular momentum relative to the rigid body center of mass is given by $J_c=I_c \omega '$, with $\omega' = d\theta /dt$, and noticing that $d x = v_x dt$ (holonomic constraint), it is readily obtained
\begin{equation}\label{eq7d}
M \ddot{x}= M a_x = M g \sin \theta - I_c \omega' \frac{d \omega'}{v_x d t} = Mg \sin \theta - \beta MR^2\frac{\omega'}{\omega' R} \alpha.
\end{equation}
Since $\alpha=a/R$, then it results the well-known equation
\begin{equation}\label{eq7e}
a_x=\frac{g \sin \theta}{(1+\beta)}.
\end{equation}

\begin{figure}
  \includegraphics[width=3.0 in]{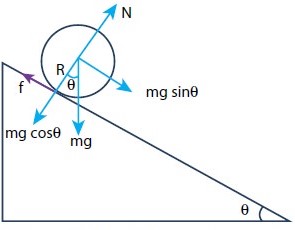}\\
  \caption{Rolling body on an inclined plane.}\label{fig1}
\end{figure}

\subsection{The fluid equation}

The {\it internal mechanical energy} term, $U^{(\alpha)}_{mec}$, appearing in Eq.~\ref{eq2} may be defined by:
\begin{equation}\label{eq2a}
U^{(\alpha)}_{mec} = m^{(\alpha)} \phi^{(\alpha)} (\mathbf{r})
+ m^{(\alpha)} \sum_{\substack{\beta=1 \\ \beta \neq \alpha}}^N \phi^{(\alpha,\beta)}.
\end{equation}
Considering Eq.~\ref{eq2a} and the definition of thermodynamic temperature, $\partial S^{(\alpha)} / \partial U^{(\alpha)} \equiv1/T^{(\alpha)}$, it follows that

\begin{eqnarray}\label{eq2aa}
\pmb{\nabla}_{r^{(\alpha)}} U^{(\alpha)} = - m^{(\alpha)} \pmb{\nabla} \phi^{(\alpha)} - \frac{\mathbf{p}^{(\alpha)}}{m^{(\alpha)}} \cdot \pmb{\nabla} \mathbf{p}^{(\alpha)} - \pmb{\nabla}_{r^{(\alpha)}} \left( \frac{J^{(\alpha)}2}{2I^{(\alpha)}} \right) 
&& \nonumber\\ - q^{(\alpha)} \pmb{\nabla} V^{(\alpha)} + q^{(\alpha)} \pmb{\nabla} (\mathbf{A}^{\alpha} \cdot \mathbf{v}^{(\alpha)}).
 \end{eqnarray}
 
Eq.~\ref{eq2aa} contains the particle's self-energy and the particle interaction energy for the gravitational and electromagnetic fields, but it may also include other terms, such as terms included in Eq.~\ref{eq2a}, representing different occurring phenomena (exemplifying energy as a bookkeeping concept).
We may recall that the entropic flux in space is a type of generalized force
$X_{\alpha}$ ~\cite{Prigogine:71,Khinchin}; therefore, it can be shown that the following equation holds:
\begin{displaymath}
T \pmb{\nabla}_{r^{(\alpha)}} \mathfrak{S}^{(\alpha)} = - q^{(\alpha)} \pmb{\nabla}_{r^{(\alpha)}} V^{(\alpha)}
+ q^{(\alpha)} \pmb{\nabla}_{r^{(\alpha)}} (\mathbf{A}^{(\alpha)} \cdot \mathbf{v}^{(\alpha)})
\end{displaymath}
\begin{equation}\label{eq2aaa}
+ m^{(\alpha)} \mathbf{v}^{(\alpha)} \cdot \pmb{\nabla} \mathbf{v}^{(\alpha)}
- \pmb{\nabla}_{r^{(\alpha)}} \left( \frac{(\pmb{J}^{(\alpha)})^2}{2 I^{(\alpha)}} - \pmb{\omega} \cdot \pmb{J}^{(\alpha)} \right),
\end{equation}
We can now write the fundamental equation of thermodynamics using the form of a spacetime differential equation:
\begin{equation}\label{eq2ab}
T \pmb{\nabla} \overline{S} + \sum_{\alpha} m^{(\alpha)} \frac{\partial \mathbf{v}^{(\alpha)}}{\partial t} = \sum_{\alpha} \pmb{\nabla} \mathbf{U}^{(\alpha)}.
\end{equation}
Taking into account the convective derivative, $d\mathbf{v}^{(\alpha)}/dt \equiv \partial \mathbf{v}^{(\alpha)}/\partial t + \mathbf{v}^{(\alpha)} \cdot \pmb{\nabla} \mathbf{v}^{(\alpha)}$, we obtain:
\begin{displaymath}
m^{(\alpha)} \frac{d \mathbf{v}^{(\alpha)}}{d t} = - T \pmb{\nabla}_{r^{(\alpha)}} \mathfrak{S}^{\alpha} - m^{(\alpha)} \pmb{\nabla} \phi^{(\alpha)} - q^{(\alpha)} \pmb{\nabla} V^{(\alpha)}
\end{displaymath}
\begin{equation}\label{eq2b}
+ q^{(\alpha)} \pmb{\nabla} (\mathbf{A}^{(\alpha)} \cdot \mathbf{v}^{(\alpha)})
- \pmb{\nabla}_{r^{(\alpha)}} \left( \frac{(\mathbf{J}^{(\alpha)})^2}{2 I^{(\alpha)}} - \pmb{\omega} \cdot \mathbf{J}^{(\alpha)} \right) + \mathbf{F}^{(\alpha)}_{ext}.
\end{equation}

For conciseness, the term $U^{(\alpha)}$ now includes all forms of energy inserted into the above Eq.~\ref{eq2}. On the right-hand side (r.h.s.), the first term must be present whenever the mechanical and thermodynamical equilibrium conditions are not fulfilled, the second term is the gravitational force, the third and fourth terms constitute the Lorentz force, the fifth term is a new term that represents the transport of angular momentum, and the last term represents other external forces that are not explicitly included but still act on the particle with index $(\alpha)$.

The present formalism was applied in a previous article~\cite{Pinheiro_04}, and therein we obtained the ponderomotive forces acting on a charged particle. For a neutral particle or body in a gravitational field, Eq.~\ref{eq2b} points to a type of extended fundamental equation of dynamics for a given species $(\alpha)$ at equilibrium and at a given point of space-time (Eulerian description):
\begin{equation}\label{eq3}
m^{(\alpha)} \frac{\partial \mathbf{v}^{(\alpha)}}{\partial t} = -m^{(\alpha)} \pmb{\nabla} \phi^{(\alpha)} - \pmb{\nabla}_{r^{(\alpha)}} \left( \frac{\mathbf{J}^{(\alpha)2}}{2 I^{(\alpha)}} - \pmb{\omega} \cdot \mathbf{J}^{(\alpha)} \right).
\end{equation}
Eq.~\ref{eq3} gains a new term because the body possesses an intrinsic angular momentum. In a non-rotating frame of reference, we set $\omega=0$, wherein we use the work-energy theorem to obtain the total mechanical energy of the system: $\mathcal{E}_{mec}=K+U+J_c^2/2I_c$. This is a common approach in classical mechanics. We are interested in the effect of a given force at a given space-time coordinate, not in its effect along the particle trajectory. It is worth noting that Eq.~\ref{eq3} was obtained through a variational procedure in contrast to the usual conservation theorem used, for example, in Ref.~\cite{Liboff,Huang}.

\vspace{1.0cm}

Included in the internal energy term is the interpressure term (see Eq.~\ref{eq2a}; note that here we consider a
homogeneous and isotropic fluid). The above described framework (see also Ref.~\cite{Pinheiro_04} for additional information) leads us to the well-known
hydrodynamic equation for a given species $(\alpha)$:
\begin{displaymath}
m^{(\alpha)} \frac{d \mathbf{v}^{(\alpha)}}{dt} =
- \pmb{\nabla}_{r^{(\alpha)}} (m^{(\alpha)} \phi^{(\alpha)})
- m^{(\alpha)} \pmb{\nabla}_{r^{(\alpha)}} \sum_{\substack{\beta=1\\ \beta \neq \alpha}}^N \phi^{(\alpha,\beta)}
\end{displaymath}
\begin{equation}\label{eq4}
- \pmb{\nabla}_{r^{(\alpha)}} \left( \frac{J^{(\alpha)2}}{2I^{(\alpha)}} - \pmb{\omega} \cdot \mathbf{J}^{(\alpha)} + T \mathfrak{S}^{(\alpha)} \right).
\end{equation}
In the r.h.s. of Eq.~\ref{eq4}, we explicitly introduce external force terms, eventually present in open systems.

Using the following correspondence from particle to fluid descriptions
\begin{equation}\label{eq5}
\sum_{\alpha} m^{(\alpha)} \rightarrow \int_V d^3 x^{'}
\rho_v(\mathbf{x'}),
\end{equation}
and an analogous relation for the electric charge
\begin{equation}\label{eq5a}
\sum_{\alpha} q^{(\alpha)} \rightarrow \int_V d^3 x^{'} \rho(\mathbf{x'}),
\end{equation}
we can rewrite Eq.~\ref{eq4} using the form of the Euler (governing)
equation:
\begin{equation}\label{eq6}
\rho_v \frac{d \mathbf{v}}{d t} = - \rho_v \pmb{\nabla}_r \phi - \pmb{\nabla}_r p - \pmb{\nabla}_r \Phi_J - \pmb{\nabla}_r f.
\end{equation}
Here, as usual, the total interparticle pressure term (e.g., Ref.~\cite{Chandrasekhar1}) is given by:
\begin{equation}\label{eq7}
p (\mathbf{r})= \sum_{\alpha} m^{(\alpha)} \sum_{\substack{\beta=1\\\beta \neq \alpha}}^N
\phi^{(\alpha,\beta)}(\mathbf{r}).
\end{equation}
To simplify, we introduce a functional integral of the form of an intrinsic angular momentum energy density (comprising the ``interaction energy term", $\pmb{\omega} \cdot \mathbf{J}$), $\Phi_J$:
\begin{equation}\label{eq7a}
\sum_{\alpha} \left[ \frac{J^{(\alpha)2}}{2I^{(\alpha)}} - \pmb{\omega} \cdot \mathbf{J}^{(\alpha)} - (\Delta F)^{(\alpha)} \right] \to \int [\Phi_J(\mathbf{x}') + f(\mathbf{x}')] d^3 x',
\end{equation}
considering that the intrinsic angular momentum density refers to a given blob of fluid (with inertial momentum $I$, a measure of the local rotation, (i.e., spin, of the fluid element), and its associated free energy (per unit volume), $f=f_0-Ts$. Eq.~\ref{eq6} also suggests that the function $S^{(\alpha)}$ (the {\it field integral of $\mathbf{r}^{(\alpha)}$}) is constant along the integral curves of the space field $\mathbf{r}^{(\alpha)}$. The gradient of the free energy, $f$, of the out-of-equilibrium state is the source of the spontaneous change from an unstable state to a more stable state while performing work. For example, a common source of free energy in a collisionless plasma is an electric current~\cite{Gary_2005}; in a magnetically confined plasma, several classes of free energy sources are available to drive instabilities, e.g., the relaxation of a non-Maxwellian, non-isotropic velocity distribution~\cite{Stacey_2005}. At this stage, it is worth noting that our procedure includes the treatment of the effect of angular momentum (through Eq.~\ref{eq2}), a necessary inclusion in a consistent theory, according to Curtiss~\cite{Curtiss 1956}.

\subsection{Application to electrodynamics}

The development of the theoretical framework in the previous sections can be easily extended to the electrodynamics, showing to be a useful pedagogical device. In addition, the method is fruitful since a new fourth term of force appears, the topological torsion term of force, of the type $q [\pmb{\omega} \times \pmb{A}]$ (to be discussed later). 

First, we may note that the mathematical identity holds:
\begin{eqnarray}\label{eq10a}
\pmb{\nabla} (\mathbf{A}^{(\alpha)} \cdot \mathbf{v}^{(\alpha)}) = && (\mathbf{A}^{(\alpha)} \cdot \pmb{\nabla})\mathbf{v}^{(\alpha)} + (\mathbf{v}^{(\alpha)} \cdot \pmb{\nabla})\mathbf{A}^{(\alpha)} + [\pmb{A}^{(\alpha)} \times \pmb{\nabla} \times \pmb{v}^{(\alpha)}] \nonumber\\
&& + [\mathbf{v}^{(\alpha)} \times \pmb{\nabla} \times \mathbf{A}^{(\alpha)}],
\end{eqnarray}
and, by means of algebraic manipulation, the following expression can be obtained:
\begin{equation}\label{eq10b}
\pmb{\nabla} \left( \mathbf{A}^{(\alpha)} \cdot \mathbf{v}^{(\alpha)} \right) = -\frac{\partial \mathbf{A}^{(\alpha)}}{\partial t} - \left[ \pmb{\omega} \times \mathbf{A}^{(\alpha)} \right] + \left[ \mathbf{v}^{(\alpha)} \times \mathbf{B}^{(\alpha)} \right].
\end{equation}
Here, $\mathbf{B}=[\pmb{\nabla} \times \mathbf{A}]$. In addition, we notice that the following equality holds:
\begin{equation}\label{eq7aaa}
(\mathbf{A}^{(\alpha)} \cdot \pmb{\nabla}) \mathbf{v}^{(\alpha)} = [\pmb{\omega} \times \mathbf{A}^{(\alpha)}],
\end{equation}
where
\begin{equation}\label{eq11a}
\mathbf{A}^{(\alpha)} = \sum_{\substack{\beta=1\\\beta \neq \alpha}} q^{(\beta)} \frac{\mathbf{v}^{(\beta)}}{r_{\alpha \beta}},
\end{equation}
denotes the vector potential actuating on the particle $(\alpha)$ due to every other particle $\beta$, and the vorticity is defined by
\begin{equation}\label{eq12}
\pmb{\Omega^{(\alpha)}} = [\pmb{\nabla}_{\mathbf{r}^{(\alpha)}} \times
\mathbf{v}^{(\alpha)}] = 2 \pmb{\omega^{(\alpha)}}.
\end{equation}

Therefore, the general equation of dynamics for a physical system (Lagrangian description) follows:
\begin{equation}\label{eq7b}
\rho \frac{d \mathbf{v}}{d t} = \rho \mathbf{E} + [\mathbf{J} \times \mathbf{B}] - \pmb{\nabla} \phi -
\pmb{\nabla} p + \rho [\mathbf{A}  \times \pmb{\omega}].
\end{equation}
The last term in the r.h.s. of the Eq.~\ref{eq7b} is a new term that represents a type of topological spin vector~\cite{Kiehn_2007}, an artifact of non-equilibrium process. We will show that the topological spin vector plays a role in plasma arcs, as well as in magnetocumulative generators~\cite{Onoochin_2003} and explosive-driven generators to produce high current and energy pulses suitable to excite high power lasers or thermonuclear fusion reactors~\cite{Altgilbers}, and suggests a new method for obtaining the helicity transport equation~\cite{Ishida:97}.

\subsubsection{Application to plasma physics: the plasma torch}

The compression of an electric current by a magnetic field, the z-pinch effect, can be studied on the basis of Eq.~\ref{eq7b}, which gives the condition for dynamic equilibria. Let us assume a typical geometry for an infinitely long axisymmetric cylindrical arc (Fig.~\ref{fig1}) with axial current density $J_z=J_z(r)$. Because the current density is assumed to be constant, Maxwell's equations in the steady state yield the azimuthal component $B_{\theta}=\mu_0 J_z r/2$ for $r \leq R$ with $R$ the outer boundary of the cylindrical arc. The vector potential is purely radial and is given by $A_z(r)=-\mu_0 R^2 J_z/4$, for $r<R$, where the Coriolis term plays no role. Based on the above conditions, we can write Eq.~\ref{eq7b} in the form:
\begin{equation}\label{eq15a}
    -J_z B_{\theta} = \frac{dp}{dr}.
\end{equation}
It follows
\begin{equation}\label{eq15b}
p(r)=\int_r^R \frac{dp}{dr} dr = \frac{1}{4}\mu_0 J_z^2 (R^2 - r^2),
\end{equation}
which is a well-known result.

The interaction between vacuum arcs and transverse magnetic fields is used in switching devices for repetitive high current interruption (see e.g. Refs.~\cite{Klajn 1999,Flurscheim}). We can instead consider a coaxial configuration with a cathode on-axis with a stabilizing magnetic induction field $\mathbf{B}$ directed along the axis of symmetry and an arc current density $\mathbf{J}$ flowing radially (and assuming a "filamentary" current with radius $R'$, $A_r=-\mu_0 R'^2 J_r/4$, with $\mu_0$ representing the permeability of the vacuum). In this case, we may apply Eq.~\ref{eq15b} and obtain the pressure differential, from the axis to the wall (at $R$):
\begin{equation}\label{eq15c}
\Delta p(r=R)=2 \pi R \left[ -J_r B_z + \mid \rho_c \mid \pi R^2 \frac{\mu_0}{4 \pi} J_r \omega \right] \Delta \theta.
\end{equation}

The conditions for retrograde or amperian rotation are the following:

\begin{subequations}\label{eq15d}
\begin{align}
    \mid \rho_c \mid > \frac{\mu_0}{4 \pi} \frac{B_z}{S} \omega ,  & \;\;\; \text{retrograde rotation} \\
    \mid \rho_c \mid < \frac{\mu_0}{4 \pi} \frac{B_z}{S} \omega ,  & \;\;\; \text{amperian rotation} .
\end{align}
\end{subequations}

%\begin{subequations}\label{eq15d}
%\begin{align}
%    \mid \rho_c \mid > \frac{\mu_0}{4 \pi} \frac{B_z}{S} \omega ,  & \;\;\; \text{retrograde rotation} \;\;\;\;\;\;\;\;\; (a) \nonumber \\
%    \mid \rho_c \mid < \frac{\mu_0}{4 \pi} \frac{B_z}{S} \omega ,  & \;\;\; \text{amperian rotation} \;\;\;\;\;\;\;\;\;\; (b) \nonumber
%\end{align}
%\end{subequations}

Here, $S$ denotes the filamentary current cross-section, with $\rho=\rho_c$. For negative charge carriers ($\rho_c=-\mid \rho_c \mid=-en_e$), we obtain an Amperian (clockwise) rotation for high magnetic fields and relatively weak arc currents. We also find a retrograde rotation for higher intensity arcs (higher $S$) and small transverse magnetic fields, which is in agreement with experimental evidence (e.g., Ref.\cite{Klajn 1999}). Here, we see the interplay between the tendencies of the energy to attain a minimum value, while the entropy attempts to attain a maximum value. From Eq.~\ref{eq15c}, we obtain an expression for the spot velocity in a transverse magnetic field:
\begin{equation}\label{eq15dd}
v=\omega R=\frac{4 \pi}{\mu_0}\frac{1}{\mid \rho_c \mid} \left( \frac{\epsilon_0 i}{2 \pi \sigma_c} + \frac{B_z R}{S} \right).
\end{equation}
Here, we have made use of the Bernoulli relation, $\Delta p=\epsilon_0 E^2/2$, and the constitutive equation, $J_r=\sigma_c E$, where $\sigma_c$ denotes the electrical conductivity of the plasma. Although Eq.~\ref{eq15dd} is not self-consistent, it shows that the force term $[\mathbf{J} \times \mathbf{B}]$ is not the only player in the process, and that the spot velocity depends linearly on the arc current (see, e.g., Ref.~\cite{Boxman 1995}). This approach shows that in out-of-equilibrium systems, a new force term occurs in a way that can suppress the Amperian force under the above referred conditions given by Eqs.~\ref{eq15d} (see also Ref.~\cite{DRJ_62}).

\begin{figure}
  \includegraphics[width=3.0 in]{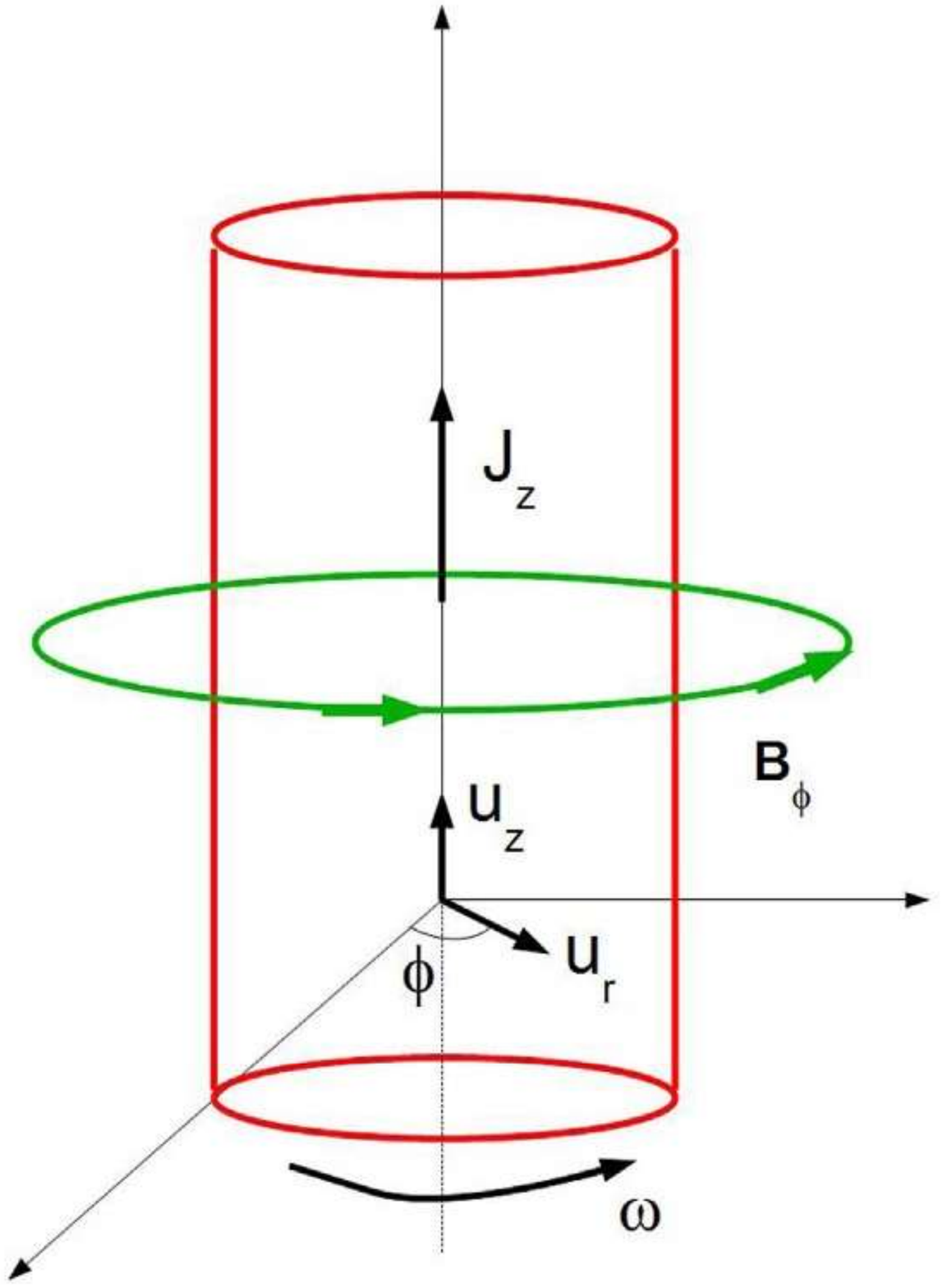}\\
  \caption{Geometry and vectorial fields in the Bennett pinch generated by an axial current $J_z$ creating a toroidal field $B_{\phi}$. If, instead, we consider a vacuum arc discharge with radial current $J_r$ and magnetic field $B_z$, we find a rotating arc with an angular velocity $\omega$.}\label{fig1}
\end{figure}

A better understanding of this phenomenon is crucial, because arc discharges are powerful generators of non-equilibrium atmospheric pressure plasmas. We compare the results predicted by Eq.~\ref{eq15dd} with experimental data available in the literature in Table 1. The calculations were done using the expression for the electrical conductivity in terms of the microscopic parameters of the plasma, $\sigma_0=e n_e \mu = \varepsilon_0 \omega_{pe}^2/\nu_c$, with the electronic mobility given by $\mu=e/m \nu_c$ ($nu_c$ denotes the electron collision frequency) because, for the majority of the data, the transverse magnetic field was below 0.1 T and is not expected to greatly influence  the plasma arc electrical conductivity~\cite{Wang_1990}.

\begin{table}
\caption{Please write your table caption here}
\label{tab:1}       % Give a unique label
\begin{tabular}{lll}
\hline\noalign{\smallskip}
first & second & third  \\
\noalign{\smallskip}\hline\noalign{\smallskip}
number & number & number \\
number & number & number \\
\noalign{\smallskip}\hline
\end{tabular}
\end{table}

\begin{table}
\caption{\label{tab:table1} Comparing Eq.~\ref{eq15dd} with experimental data for low pressure and atmospheric DC discharges. Electric current $i=60$ A.}
%\begin{ruledtabular}
\begin{tabular}{lll}
\hline\noalign{\smallskip}
                              &  Low pressure                 & Atmospheric pressure  \\
\hline\noalign{\smallskip}
$T_e (eV)$                    &  1.0 \footnotemark[1]                       & 0.87 \footnotemark[1]\\
Electron density (cm$^{-3}$)  & $n_e=2.5 \times 10^{12}$ \footnotemark[6] & $n_e = 2 \times 10^{16}$ \footnotemark[2]\\
collision frequency (s$^{-1}$)& $\nu_c=6.98 \times 10^7$                  & $\nu_c=7.2 \times 10^{11}$ \footnotemark[3]\\
Plasma frequency (s$^{-1}$)   & $\omega_{pe}=8.94 \times 10^{10}$         & $\omega_{pe}=8 \times 10^{12}$ \footnotemark[4]\\
Average Speed (m$/$s)         & 3.0 ($\sim 2$) \footnotemark[5]              & $10^{-3} (5\times 10^{-3})$ \footnotemark[5]\\
\end{tabular}
%\end{ruledtabular}

\footnotetext[1]{For laboratory discharges, the Coulomb logarithm is $\ln \Lambda \sim 10$, see Ref.\cite{Chen}, for electron temperatures of the order of $T_e \approx 10000$ K~\cite{Tu_2007}}
\footnotetext[2]{Ref.~\cite{Tu_2007}}
\footnotetext[3]{The frequency of collision was calculated using the standard expression $\nu_c = 2.91 \times 10^{-6} n_e \ln \Lambda T_e^{-3/2}$, Ref.~\cite{Chen}].}
\footnotetext[4]{We use $\omega_{pe}=5.65 \times 10^4 \sqrt{n_e}$, (see Ref.\cite{Chen}).}
\footnotetext[5]{In parenthesis are the experimental data collected for atmospheric pressure, from Ref.~\cite{DRJ_62}; for low pressure, see Ref.~\cite{Zhu_2013}.}
\footnotetext[6]{Data interpolated from Ref.~\cite{Schram_1997}, assuming $T_e=0.8$ eV.}
\end{table}

\subsection{Poynting's Theorem}

We must now proceed from the notion of the localization of electromagnetic energy in space, and generalize the concept to other forms of energy: $\mathfrak{I}$,kinetic energy; $\mathfrak{P}_g$, total gravitational energy; $\mathfrak{P}_v$, total electrostatic energy; and, $\mathfrak{P}_{rot}$, the total rotational energy.

From Eq.~\ref{eq4} it can be obtained an energy conservative equation:
\begin{displaymath}
\frac{\partial }{\partial t}(\mathfrak{I}+ \mathfrak{P}_g +
\mathfrak{P}_v + \mathfrak{P}_{rot}) = -\frac{1}{2} \sum_{\alpha}
\pmb{\nabla}_{\mathbf{r}^{(\alpha)}}  \cdot \left\{ \mathbf{v}^{(\alpha)}
\left[ -\frac{|\mathbf{J}^{(\alpha)}|^2}{2 I^{(\alpha)}}
+ (\pmb{\omega} \cdot \mathbf{J}^{(\alpha)})
- T \mathfrak{S}^{(\alpha)} \right] \right\}
\end{displaymath}
\begin{equation}\label{eq27}
+ \frac{1}{2} \sum_{\alpha} \left[ -\frac{|\mathbf{J}^{(\alpha)}|^2}{2
I^{(\alpha)}} + (\pmb{\omega} \cdot \mathbf{J}^{(\alpha)})
- T \mathfrak{S}^{(\alpha)} \right] (\pmb{\nabla}_{\mathbf{r}^{(\alpha)}} \cdot \mathbf{v}^{(\alpha)}).
\end{equation}
Finally, we can transform Eq.~\ref{eq27} into a version of {\it
Poynting's Theorem for rotating fluids}:
\begin{equation}\label{eq28}
-\frac{\partial }{\partial t}(\mathcal{U}) = \pmb{\nabla} \cdot \mathbf{S} +
\mathcal{P}',
\end{equation}
on which we define a type of {\it Poynting vector for rotational fluids}, which gives the rate of rotational energy flow:
\begin{equation}\label{eq29}
\mathbf{S}=\frac{1}{2} \sum_{\alpha} \mathbf{v}^{(\alpha)}\left[
-\frac{|\mathbf{J}^{(\alpha)}|^2}{2 I^{(\alpha)}} +
(\pmb{\omega}^{(\alpha)} \cdot \mathbf{J}^{(\alpha)}) -
T \mathfrak{S}^{(\alpha)} \right].
\end{equation}
The {\it power input} driving the process (source$/$sink term) is given by:
\begin{equation}\label{eq30}
\mathcal{P}' \equiv \frac{1}{2} \sum_{\alpha} (\pmb{\nabla} \cdot \mathbf{v}^{(\alpha)}) \left[
\frac{|\mathbf{J}^{(\alpha)}|^2}{2 I^{(\alpha)}} - (\pmb{\omega}
\cdot \mathbf{J}^{(\alpha)})
+ T \mathfrak{S}^{(\alpha)} \right],
\end{equation}
The total energy is defined by summing up the different contributions:
\begin{equation}\label{eq30a}
\mathcal{U} = \mathfrak{I} + \mathfrak{P}_g + \mathfrak{P}_v + \mathfrak{P}_{rot}.
\end{equation}

Here, the term $T \mathfrak{S}^{(\alpha)}$ represents the thermal energy associated with the species $(\alpha)$ that is
equal to $-\Delta F$, the free energy of the physical system. For
a system in contact with a reservoir at constant temperature this is
the maximum amount of work extractable from the system; the free energy
tends to decrease for a system in thermal contact with a heat
reservoir. 

\subsection{Driving energy of a rotating system}

In particular, notice that when
the angular velocity, $\omega$, is multiplied by Eq.~\ref{eq4}, the driving power is obtained. It is worth noting that the presence of the term $\pmb{\omega} \cdot \mathbf{J}$, which plays an analogous role to the slip in electrical induction motors, that is, the lag between the rotor speed and the magnetic field's speed, is provided by the stator's rotational speed. Furthermore, we see that the power input depends on the fluid compressibility $\pmb{\nabla} \cdot \mathbf{v}$. This means that compressibility is a factor that determines the amount of transported angular momentum through the stress-tensor $\tau_{ij}$ and may be responsible for a new driving mechanism in addition to the well-known MRI. The driving energy of the rotating system can be expressed in the form:
\begin{equation}\label{eq30b}
\mathcal{E}_{driv} = \frac{J^2}{2I_p} - (\pmb{\omega} \cdot \mathbf{J}) -\Delta F.
\end{equation}

Next, we will discuss several examples illustrating the application of the variational method.

\subsection{Transport of angular momentum in a hurricane}

A hurricane is a natural airborne structure that converts its kinetic and potential energy by means of the transport of angular momentum from the inner core to the outer regions, conveyed either directly by moving matter, or by non-material stresses such as those exerted by electric or magnetic fields~\cite{Valdis}. We may apply Eq.~\ref{eq30} to this specific problem, assuming that all of the mechanical and thermal energy is converted into electromagnetic energy $U_e$, to obtain:
\begin{equation}\label{eq31}
\mathcal{P}'=\omega \left( \frac{J^2}{2I_p} - \pmb{\omega} \cdot \mathbf{J} - \Delta F \right) = -\frac{\partial U_e}{\partial t}
\end{equation}
or
\begin{equation}\label{eq31a}
\mathcal{P}' = \omega \mathcal{E}_{driv}.
\end{equation}
Let us consider the case of a hurricane in an axisymmetric configuration, with $J=\omega I$. We can safely assume that $\omega^2 I^2/I_p \gg \omega^2 I + \Delta F$. We can now envision a simple model of a hurricane with a total mass, $M$, and radius, $R$, approximated as a solid cylinder with $I=MR^2/2$. Hence, the total power driving the hurricane is given by
\begin{equation}\label{eq31b}
\mathcal{P}' = \frac{1}{8} \frac{\omega^3}{2} \frac{M^2R^4}{I_p} \propto \omega^3 \frac{M^2R^4}{I_p},
\end{equation}
or, as a function of the fluid density $\rho$:
\begin{equation}\label{eq31c}
\mathcal{P}' = \frac{\pi^2}{4} \omega^3 \frac{R^6 L^2}{I_p} \rho^2.
\end{equation}
Our result shows the same type of dependency that was demonstrated by Chow $\&$ Chey~\cite{Chow}, and, in particular, it shows that the intrinsic inertial momentum of the particles constituting the fluid plays a substantial role.

\subsection{Periodic radiative heating of the Earth's atmosphere}

It has been experimentally shown~\cite{Schubert_69} that periodic radiative heating of the Earth's atmosphere transmits angular momentum to it as a result of the Earth-atmosphere coupling through frictional interactions ~\cite{Chandrasekar}. The images sent by the ESA's Venus Express confirms this fact on Venus (Earth's planetary twin) based on the presence of a ``double-eye" atmospheric vortex at the planet's south pole and the presence of high velocity winds whirling westwards around the planet, which is characterized by a four-day period.

\begin{figure}
    \centering
    \includegraphics[width=3 in]{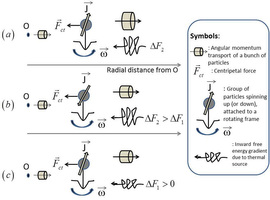}
    \caption{A group of particles spinning about their axes and revolving around a common axis, Oz, subject to a centripetal force. Three situations typically occur. (a) An outward transport of angular momentum occurs with a larger gradient, while free energy flows to the center of the field. (b) If the inward free energy gradient is dominant relative to the angular momentum gradient, a reversal of the particle' angular momentum may occur. (c) If the angular momentum gradient is of the same order of magnitude as in case (b) but still dominant relative to the inward free energy gradient, the particles may continue spinning in the same direction.}
    \label{fig3}
\end{figure}

Schubert and Withehead's~\cite{Schubert_69} conducted an experiment with the purpose of providing an explanation for the high wind velocities during apparent cloud formation in the upper atmosphere of Venus. In this experiment, a Bunsen flame rotating below a cylindrical annulus filled with liquid mercury induced the rotation of the liquid mercury in a direction counter to that of the rotating flame. The speed of the flame was 1 mm$/$s and the temperature of the mercury increased from room temperature at a rate of approximately 3$^0$C per minute. After 5 minutes, a steady-state flow was established, with the mercury rotating in the counter-direction of the flame, with a speed of approximately 4 mm$/$s. If we consider the liquid to be a spinning body, we can estimate that
\begin{equation}\label{eq31d}
\frac{d}{dr} \left( \frac{I_p \omega'^2}{2} - \frac{I_p \omega^2}{2} + T \mathfrak{S} \right) > 0 \Rightarrow I_p \Delta(\omega' + \omega)(\omega' - \omega)+\Delta (T \mathfrak{S}) > 0.
\end{equation}
Hence, the following result is obtained:
%\begin{equation}\label{eq32}
%\omega' \Delta \omega \gtrsim -\frac{1}{2}\frac{\Delta (T \mathfrak{S})}{I_p}.
%\end{equation}
\begin{equation}\label{eq32a}
\omega' \gtrsim \pm \omega \sqrt{1-\frac{\Delta F}{2I_p \omega^2}},
\end{equation}
and, in the limit $\omega \approx \omega'$,
\begin{equation}\label{eq32}
\omega' \Delta \omega \gtrsim -\frac{1}{2}\frac{\Delta T \mathfrak{S}}{I_p}.
\end{equation}
Here, $\omega$ is the angular speed imposed on the system (heat source), $\omega'$ is the mass flow induced angular speed due to a sustained source of energy; $\Delta \omega \equiv \omega' - \omega$. We use the following tabulated data: $\mathfrak{S}=16.6$ J.K$^{-1}$mol$^{-1}$ for mercury at temperature of the experiment and $\rho \approx 13.6 \times 10^{-3}$ kg$/$m$^3$. We also consider that the volume of mercury is contained in the channel of the experimental apparatus forming a rim with an average radius of $R=30$ cm. Using Eq.~\ref{eq32}, we estimate that after one minute, the speed must be approximately 3.7 mm$/$s. It is clear that the sense of rotation and the speed of the wind depend on the latent heat stored in the planetary atmosphere and the temperature difference between the boundaries (through $\Delta T$). Eq.~\ref{eq32a} is consistent with the results reported by~\cite{Douglas_1972}, where a maximum mean surface flow is observed corresponding to the angular velocity of the heat source when the convective velocity (in our example, $v_{stam}=\sqrt{\Delta F/2 M_{\mu}}$, with $M_{\mu}$ denoting the molar mass) is attained. Note that, in fact, our $v_{stam}$ represents the limiting speed for the transport of angular momentum. For mercury, this velocity is $v_{stam} \approx 24$ m$/$s, with $M_{\mu}=200.59$ g$/$mol in the conditions of the moving flame experiment. The existence of a limit to the amplification of the angular speed was also suggested in Ref.~\cite{Stern_1971}, which demonstrated the effect of a heating or cooling source in the momentum equilibration. Three possible cases for the clockwise or counter-clockwise motion of a fluid in a rotation frame are observed in Fig.~\ref{fig3}.

Eq.~\ref{eq27} is not reversible in time due to the presence of the ``hyperfine structure" $(\pmb{\omega} \vdot \mathbf{J})$ and the free energy term $F$. It was shown in Ref.~\cite{Pinheiro_2013} that the power equation can be written under the form of a {\it Poynting's Theorem} for rotating fluids or extended bodies:
\begin{equation}\label{eq28}
-\frac{\partial }{\partial t}(\mathcal{U}) = \pmb{\nabla} \cdot \mathbf{S} +
\mathcal{P}',
\end{equation}
provide we first define the following Poynting vector for rotational fluids and specializing for a single specie, which gives the rate of rotational energy flow:
\begin{equation}\label{eq29}
\mathbf{S} \equiv \frac{1}{2} \left[
-\frac{|\mathbf{J}|^2}{2 I} +
(\pmb{\omega} \cdot \mathbf{J}) +F \right] \mathbf{v}.
\end{equation}
The {\it power input} driving the process (source$/$sink term) is given by:
\begin{equation}\label{eq30}
\mathcal{P}' \equiv \frac{1}{2}  (\pmb{\nabla} \cdot \mathbf{v}) \left[
\frac{|\mathbf{J}|^2}{2 I} - (\pmb{\omega}
\cdot \mathbf{J}) - F \right],
\end{equation}
and the total energy is defined by summing up the different contributions:
\begin{equation}\label{eq30a}
\mathcal{U} = \mathfrak{I} + \mathfrak{P}_g + \mathfrak{P}_v + \mathfrak{P}_{rot}.
\end{equation}
Here, $\mathfrak{I}$ is the total kinetic energy, $\mathfrak{P}_g$ is the gravitational energy, $\mathfrak{P}_v$ is the total electrostatic energy, and $\mathfrak{P}_{rot}$ is the total rotational energy (see Ref.~\cite{Pinheiro_2013} for details).
Eq.~\ref{eq28} shows that it is vortical motion and$/$or free energy processes that are at the root of non-equilibrium processes and dissipation.

\subsection{The Flyby Anomaly of the spacecrafts}

\begin{figure}
    \centering
    \includegraphics[width=3.5 in]{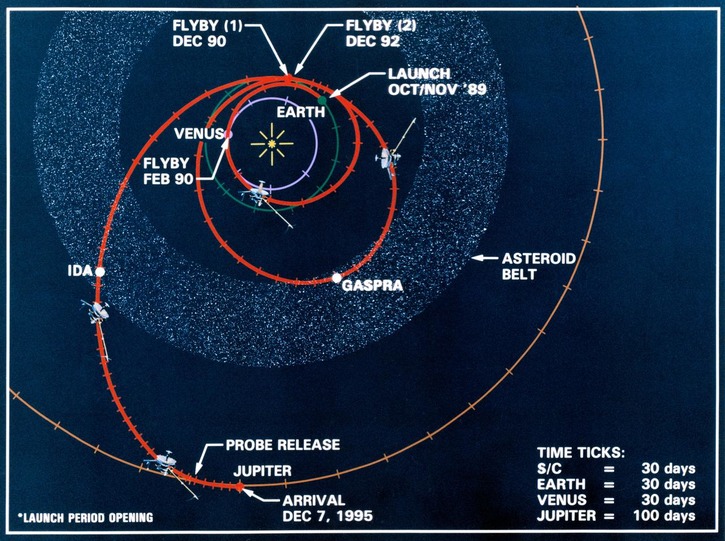}
    \caption{Trajectory drawing of the Galileo spacecraft's launch from low Earth orbit including Venus (February 1990), two Earth flybies (December 1990 and December 1992), and the asteroids Gaspra and Ida in the asteroid belt. Courtesy: NASA.}
    \label{fig:galileo}
\end{figure}

The flyby anomaly is still an unsolved problem in astrodynamics discovered by a research team at the Jet Propulsion Laboratory, lead by John Anderson. The first observation appeared when the trajectory of the Galileo spacecraft when it approached planet Earth on December 8th, 1990, followed by means of Doppler data, showed a difference on its ingoing and outgoing asymptotic velocities, differing in 3.92 mm$/$s in what was expected theoretically, based on the gravitational orthodox theory (Fig.~\ref{fig:galileo}).

Now, we attempt here a simple explanation to the Juno flyby that registered unexpectedly no anomaly.

The equation to solve is
\begin{equation}\label{eq34}
\mathbf{\ddot{r}}=-\pmb{\nabla} \phi + \mathbf{f}_{TTC}.
\end{equation}
We will use the cylindrical coordinate frame $\mathcal{F}_0$ (see Ref.~\cite{Ruiter}) with basis vectors $(\mathbf{x}_o, \mathbf{y}_o, \mathbf{z}_o)$. In this coordinate frame, the position and velocity are given by
\begin{eqnarray}
  \mathbf{r} &=& r\mathbf{x}_o \\
  \mathbf{v} &=& \dot{r}\mathbf{x}_o+r \dot{\theta}\mathbf{y}_o
\end{eqnarray}
and the angular momentum is given by
\begin{equation}\label{eq35}
\mathbf{h}=h\mathbf{z}_o=r^2 \dot{\theta}\mathbf{z}_o.
\end{equation}
The perturbing force in $\mathcal{F}_o$ coordinates is given by
\begin{equation}\label{eq36}
\mathbf{f}_{TTC}=f_r \mathbf{x}_o+f_{\theta}\mathbf{y}_o+f_z\mathbf{z}_o.
\end{equation}

\begin{figure}
  \centering
  \includegraphics[width=3.25 in]{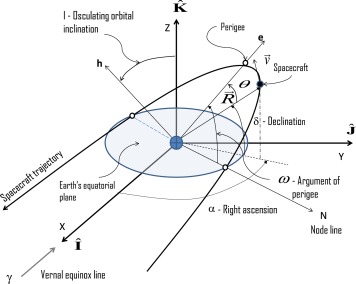}\\
  \caption{Spacecraft trajectory in flyby trajectory. Orbital parameters.}\label{Fig3}
\end{figure}

\begin{figure}
  \centering
  \includegraphics[width=3.5 in]{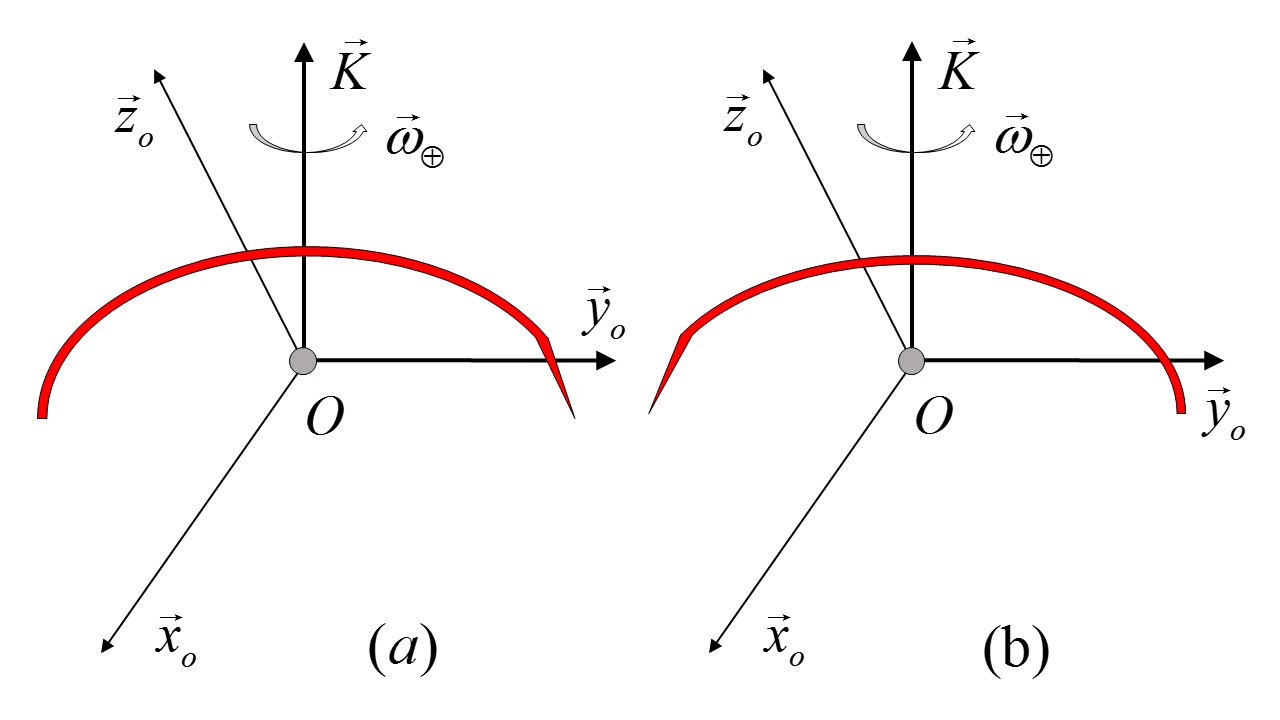}\\
  \caption{Spacecraft trajectory in retrograde (a) and posigrade (b) direction.}\label{Fig3}
\end{figure}

\subsubsection{Spacecraft in posigrade direction}~\index{Spacecraft in posigrade direction}

\begin{eqnarray}
  v_{apx} &=& V_P - v_{\infty} \cos (\omega \mp \theta) \\
  v_{apy} &=& v_{\infty} \sin (\omega \mp \theta) \\
  v_{apz} &=& 0.
\end{eqnarray}
with
\begin{equation}\label{eq37}
\mathbf{n}^{'}=\sin(\omega \mp \theta) \mathbf{y}_o + \cos (\omega \mp \theta) \mathbf{x}_o
\end{equation}
and therefore, the term $(\mathbf{v} \cdot \mathbf{n}')$ appearing in Eq.~\ref{eq11} transforms into (we solve now in the frame of the planet, hence $V_P=0$)
%\begin{widetext}
\begin{eqnarray}
  (\mathbf{v} \cdot \mathbf{n}^{'}) &=& v_{apx} \cos (\omega \mp \theta) + v_{apy} \sin(\omega \mp \theta) \\
   &=& v_{\infty} \cos^2 (\omega \mp \theta) - v_{\infty} \sin^2 (\omega \mp \theta) \\
    &=&  v_{\infty} \cos 2(\omega \mp \theta).
\end{eqnarray}
%\end{widetext}
As we will see next, Eq.~\ref{eq34} is determinant to the issue and, in particular, the dot product $(\mathbf{v} \cdot \mathbf{n}$).

To solve the set of equations ~\ref{eq34} is not our intention, we just want to solve Eq. in the region of the flyby, assuming that $v_{x_0} \approx 0$ and $R=R_{\oplus}$, and $\frac{v_{x_0} v_{\infty}}{v} \sim v_{\theta}$. The Eq. can be written in the form
\begin{equation}\label{eq38}
\frac{d v_{\theta}}{dt}=\frac{2 \omega_{\oplus} R_{\oplus}}{c}\frac{v_{x_0} c}{R_{\oplus}}\sin I + \frac{2 \omega_{\oplus} R_{\oplus}}{c}\frac{v_{x_0} v_{\infty}}{R_{\oplus}}\sin I
\end{equation}
Here, $K \equiv \frac{2 \omega_{\oplus}}{R_{\oplus}}$, or
\begin{equation}\label{4law}
\frac{d v_{\theta}}{dt}=K \frac{v_{x_0}c}{R_{\oplus}}+K\frac{v_{x_0}v_{\infty}}{R_{\oplus}}\sin I
\end{equation}
Eq.~\ref{4law} shows two forces acting together on the spacecraft. The first term is nearly zero, according to our hypothesis of nearly circular trajectory, while the second term of force actuates as a modulating force and contributes more significantly. We use the approximation of a nearly circular trajectory during the flyby, $dt=R_{\oplus}d\theta/v$.
The equation of motion for the azimuthal velocity becomes
\begin{equation}\label{eq39}
\frac{dv_{\theta}}{d \theta}= \left( \frac{K c v_r}{v} \sin I + \frac{K v_{\infty} v_r}{c} \cos 2(\omega \mp \theta) \sin I \right) \cos (\omega \mp \theta).
\end{equation}
The mathematical solution is
%\begin{widetext}
\begin{equation}\label{eq40}
v_{\theta}= e^{(c_1 \sin \theta + c_2 \sin (3 \theta)} +
e^{(c_1 \sin (\theta) + c_2 \sin (3 \theta)} \int_1^{\theta} e^{(-c_1 \sin (\theta^{'})-
       c_2 \sin (3 \theta^{'}))} \cos \theta^{'}) d \theta^{'}.
\end{equation}
%\end{widetext}
The chosen constants are: $c_1=0.07$ and $c_2=0.0233$, that give the best approach to the NEAR spacecraft velocity anomaly.
The integral make the approach to $\sin \theta)$. Fig. shows the velocity anomaly in the posigrade direction. There is an acceleration during half-way but a deceleration that follows, giving an overall null effect. There

\begin{figure}
 \centering
\includegraphics[width=3 in]{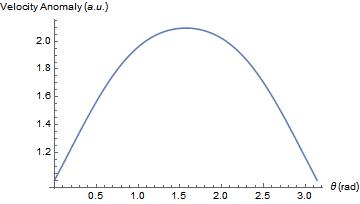}\\
\caption{Azimuthal velocity vs. $\theta$ in the posigrade direction.}\label{Fig3}
\end{figure}

\subsubsection{Spacecraft in Retrograde direction}~\index{Spacecraft in retrograde direction}

To treat this case, we need the additional components of the velocity:
\begin{eqnarray}
  v_{apx} &=& V_P + v_{\infty} \cos (\omega \mp \theta) \\
  v_{apy} &=& -v_{\infty} \sin (\omega \mp \theta) \\
  v_{apz} &=& 0.
\end{eqnarray}
and therefore,
%\begin{widetext}
\begin{eqnarray}
  (\mathbf{v} \cdot \mathbf{n}^{'}) &=& v_{apx} \cos (\omega \mp \theta) + v_{apy} \sin(\omega \mp \theta) \\
   &=& V_P \cos (\omega \mp \theta) - v_{\infty} \cos^2 (\omega \mp \theta) + v_{\infty} \sin^2(\omega \mp \theta) \\
    &=&  V_P \cos (\omega \mp \theta) + v_{\infty} \cos 2(\omega \mp \theta)
\end{eqnarray}
%\end{widetext}

\begin{figure}
  \centering
  \includegraphics[width=6 in]{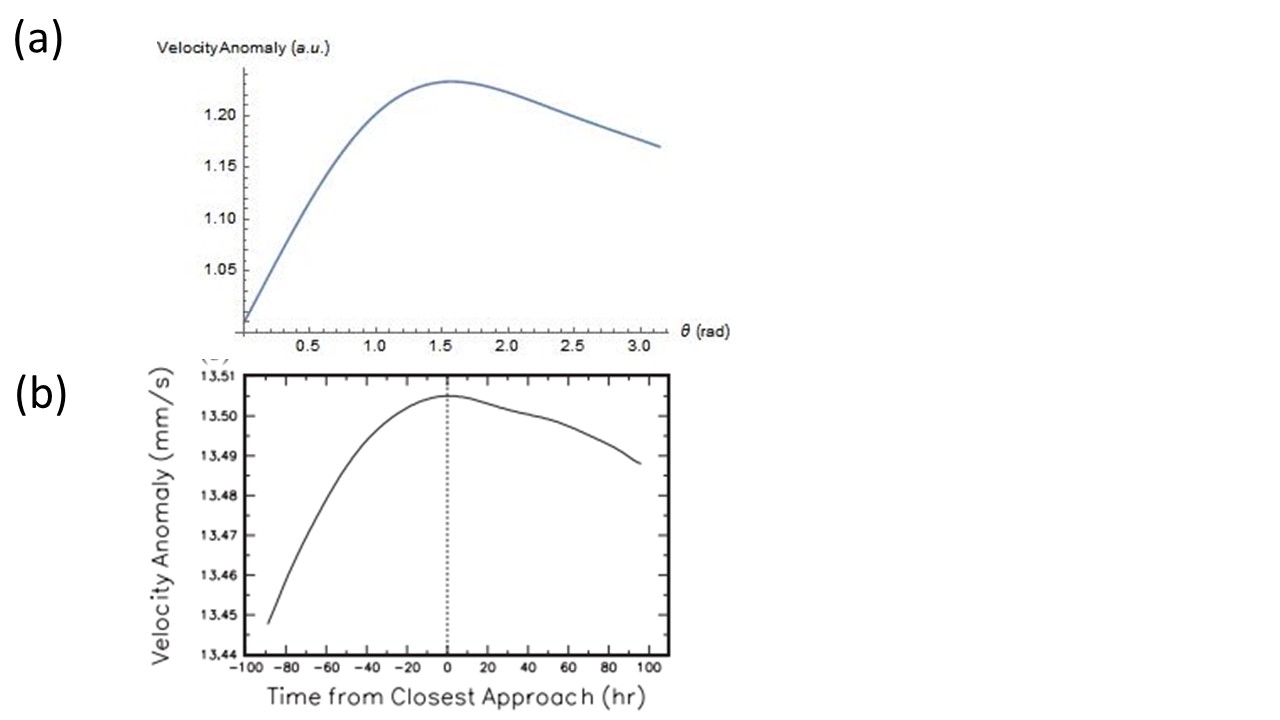}\\
  \caption{Anomalous velocity vs. $\theta$ in retrograde direction. (a) Theoretical prediction; (b) Experimental data from NEAR spacecraft, see Ref.}\label{Fig4}
\end{figure}

In this case, the equation of azimuthal motion is
%\begin{widetext}
\begin{equation}\label{eq41}
\frac{d v_{\theta}}{d t}= 2 \omega_{\oplus} v_{\infty} \cos (\omega \mp \theta)\sin I + 2 \omega_{\oplus} \frac{v_{\infty}^2}{c} \cos (\omega \mp \theta) \sin I.
\end{equation}
%\end{widetext}
Now, we may expect that $v_{\infty} \sim v_{\theta}$ and within the range of this assumption the above Eq. becomes
\begin{equation}\label{eq42}
\frac{d v_{\theta}}{d \theta}= K \frac{c}{R_{\oplus}} v_{\infty} \cos (\omega \mp \theta)\sin I  + \frac{Kv_{\infty}^2}{R_{\oplus}} \cos (\omega \mp \theta)^2 \sin I.
\end{equation}
We will make the approach and rewrite the Eq. in the new form
\begin{equation}\label{eq43}
\frac{dv_{\theta}}{d \theta}= \left( \frac{K c v_r}{v} \sin I + \frac{K v_{\infty} v_r}{c} \cos (\omega \mp \theta) \sin I \right) \cos (\omega \mp \theta).
\end{equation}
We have search the values that give the best similar shape in comparison with the radar data obtained with NEAR spacecraft, and the best guess are, approximately,
\begin{eqnarray}
  \frac{K v_{\infty} v_r}{c} &\sim & 0.14 v_{\theta} \\
  \frac{K c v_r}{v_{\theta}} &\sim & 1.
\end{eqnarray}
Considering $v_{\infty} \sim 6 \times 10^3$ km$/$s, we obtain $v_r \sim 7$ km$/$s and $v_{\theta} \sim 1$ km$/$s, values that are in the order of the expected magnitude.
The mathematical solution is given by
\begin{equation}\label{eq44}
v_{\theta} = \exp [c_3 (\omega \mp \theta) + c_4 \sin (\omega \mp \theta)](1+ c_5 \sin (\omega \mp \theta)),
\end{equation}
with $c_3=0.05$, $c_4=0.025$ and $c_5=0.14$. We may notice that the main difference between the two flyby approaches is the direct dependency on the angle $\omega \mp \theta$ in the retrograde case, which implies a bump in the shape of the figure and an asymmetry on the effect of the TTC.

Fig.~\ref{Fig4} plots the solution, comparing with the experimental data (see Fig.3(b) in Ref.~\cite{Anderson_2008}).

\subsection{Additional remarks on the variational problem according to the proposed reformulation}~\label{vario}

In 1686, Newton solved the problem of determining the shape of a rotationally symmetric body of least resistance, but the beginning of the calculus of variations is set in the year 1696 with the formulation of the brachistochrone problem by Johann Bernoulli in the {\it Acta Eruditorum}. Without recurring to minimum-time assumptions, or Euler-Lagrange equations, the basic idea for the application of the concepts of energy and entropy, closely linked to the fundamental equations of thermodynamics, create a unified description of mechanics, in particular yielding a wider frame for variational problems, such as the brachistochrone for uniform (e.g., gravitational) field.

\begin{figure}
  \centering
  \includegraphics[width=3.0 in,angle=-90 ]{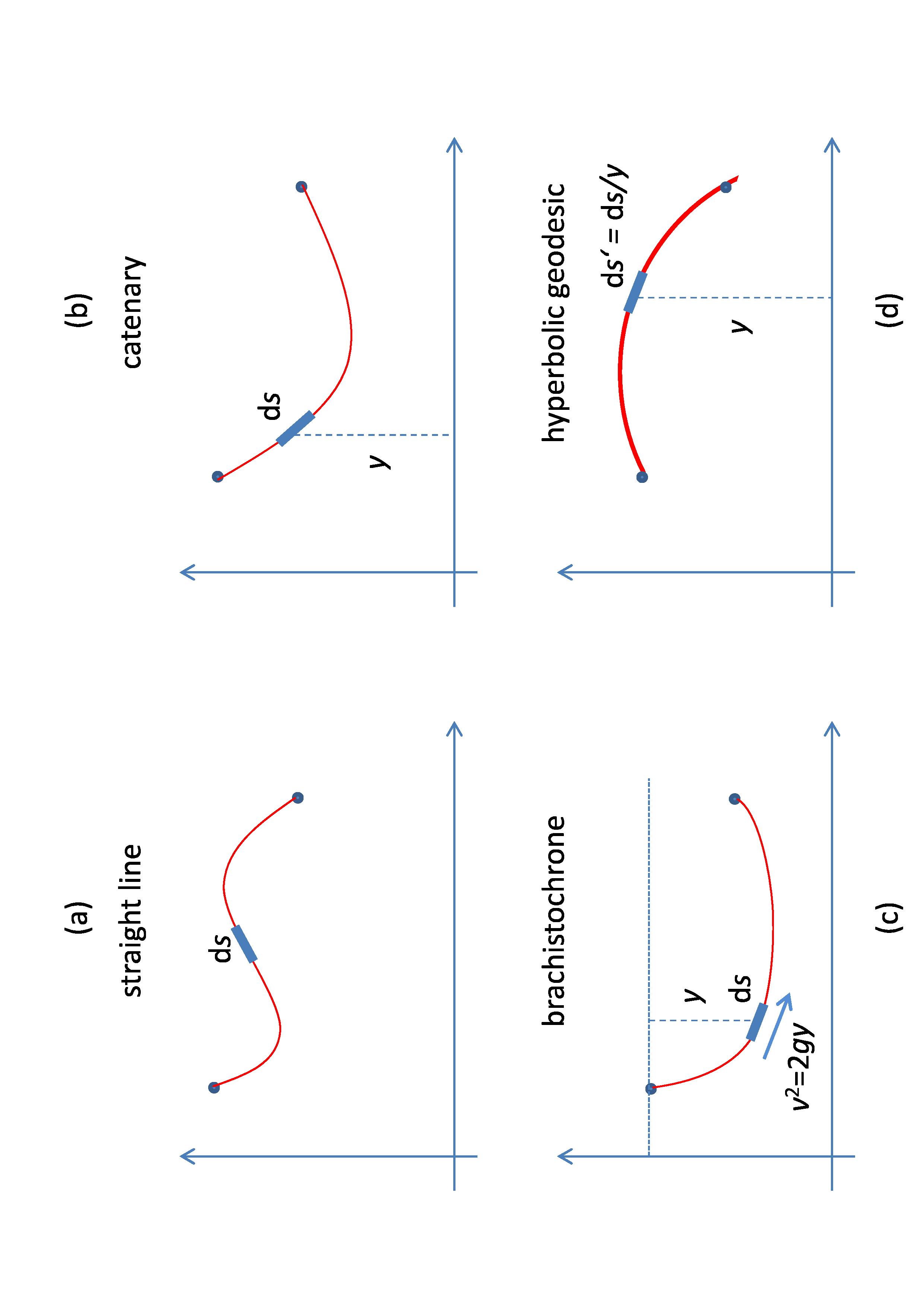}\\
  \caption{The four optimization problems: (a) finding the shortest path in Euclidean
geometry, (b) the shape of a hanging chain, (c) the path of fastest descent,
and (d) the shortest path in hyperbolic geometry. Courtesy from: Raul Rojas, Freie Universit\"{a}t Berlin, arXiv:1401.2660v1}\label{fig2}
\end{figure}

Following along our research programme, let us consider a Cartesian reference frame and a particle with mass $m$ accelerated by gravity from one point to another with no friction and solve the problem using Eq.~\ref{eq3a}. First, notice that the Cartesian frame is not rotating, hence, $\omega=0$. Noting that $\Omega$ is independent from coordinates $(x,y)$, we may assume $x(0)=-R$, $\frac{dx}{dt}=R\Omega t$ and $\frac{dy}{dt}(0)=0$, and $y(0)=R$. It gives

\begin{eqnarray}
  \frac{d^2 y}{dt^2} &=& -g-y\Omega^2 \\
  \frac{d^2 x}{dt^2} &=&   -x\Omega^2
\end{eqnarray}
since, $\nabla_y (\frac{1}{2}mr^2 \Omega^2)=my\Omega^2$ (with similar result for $x$ component), and plugging the identity $dr=\sqrt{dx^2+dy^2}$. The solutions of the above system of equations are the well-known equations of a cycloid:
\begin{eqnarray}
  x(t) &=& R(\Omega t - \sin \Omega t) \\
  y(t) &=& -\frac{g}{\Omega^2}(1-\cos \Omega t) + R\cos \Omega t.
\end{eqnarray}
The cycloid equations were straightforwardly obtained because the fundamental equation of dynamics, Eq.~\ref{eq3a}, retains consistently the two extremal principles of nature, the tendency to minimize energy and maximize entropy. There is no need to make further hypothesis.

The standard solution of one of the earliest problems in the calculus of variations starts with the calculation of the the time to travel from a point $P_1$ to another point $P_2$:
\begin{equation}\label{}
t_{12}=\int_{P_1}^{P_2}\frac{ds}{v},
\end{equation}
obtaining the speed at any point by using the equation of conservation of energy, $v=\sqrt{2gy}$. Plugging the arc length $ds=\sqrt{1+y'^2}dx$, the integral of $t_{12}$ is converted into
\begin{equation}\label{}
t_{12}=\int_{P_1}^{P_2} \sqrt{\frac{1+y'^2}{2gy}}dx=\int_{P_1}^{P_2} f(y,y')dx.
\end{equation}
Normally, the Euler-Lagrange differential equation is used, and, after some maths, another differential equation is obtained
\begin{equation}\label{}
\left[1+(\frac{dy}{dx})^2 \right]y=k^2,
\end{equation}
where $k^2$ is a positive constant. The solution is given by the parametric equations, Eqs.49-50. Our foundational project aims at finding a coherent and possibly true representation of what is physically possible without involving extraneous idealizations or empirical generalizations. In this sense, our formulation seems more natural than the Lagrangian formulation, since it represents a new synthesis of two different propensities in nature (provided by energy and entropy).

\section{Direct conversion of entropy into forces}

\subsection{The case of entropic forces}

Emergent phenomena in nature gives rise to surprising order. In particular, it seems that more complex network of interactions between elements result in fewer elements needed for the emergence of phenomena. This conception was adopted by Verlinde~\cite{Verlinde} (although, much earlier, a more in-depth proposal was advanced, and disregarded~\cite{Pinheiro_02}), and disproved~\cite{Kobakhidze}. Instead, we should look to the concept of entropic forces consistently included in the present formulation. Considering constant temperature, Eq. leads basically to $m \mathbf{a}=m\mathbf{g}-T\pmb{\nabla}S$. But we should remind that irreversible transformations all occur over positive entropy transformations, $\frac{\partial S}{\partial z}>0$, where $z$ can be a distance, temperature, composition, or volume, for example. Considering that the gravitational force is conservative, this is one of the crucial points against the conception of emergent force.

If a macroscopic system is subject to an entropic force identified as the classical force under thermal equilibrium, it is legitimate to revert the argument and examine whether some microstates provide support for the existence of referenced kinds of forces. This question is illustrate subsequently.

Imagine a particle describing a circular motion at a distance $r$
away from the origin of a central force. The number of
configurations in space associated with $r$ is $\Omega=4 \pi r^2$. The entropic force results to be
\begin{equation}\label{}
f_r = \frac{T}{2} (\vec{\nabla S})_r = \frac{k_B T}{r}.
\end{equation}
Taking into account the Equipartition Theorem, $\frac{1}{2} k_BT=
\frac{1}{2} m v_{r}^2$, when thermal equilibrium prevails. Hence,
the known expression for centrifugal force is found
\begin{equation}\label{elas1}
  f_r = m \omega^2 r.
\end{equation}

Notice that the term of mass was introduced via Equipartition
theorem. From this the expression for the force becomes
(unfortunately, the symbol for period and absolute temperature are
identical)
\begin{equation}\label{}
f_r = m \left( \frac{4 \pi^2}{T^2} \right) r = m \frac{4 \pi^2
r^3}{T^2} \frac{1}{r^2}.
\end{equation}
Recall that according to Kepler law, all the planets on the solar
system have the same ratio $\frac{r^3}{T^2}=K$:
\begin{equation}\label{}
f_r = 4 \pi^2 K \frac{m}{r^2}.
\end{equation}
Instead, Newton wrote the above equation on the form
\begin{equation}\label{Newton}
f_r = \left( \frac{4 \pi^2 K}{M_S} \right) \frac{M_S m}{r^2}=G
\frac{M_S m}{r^2}
\end{equation}
defining $G=\frac{4 \pi^2 K}{M_S}$ as the gravitational constant.
Therefore, the gravitational force is retrieved, but in the
present model on the ground of a different viewpoint, giving a new evidence to an old law of physics.

The arguments used above demand further development. First, the number of microstates $\Omega$ counts the number of times points on a volume of space in an higher space is projected on a boundary, a surface $\mathbbm{S}^2$, forming a two-dimensional information structure, suggesting a projection of information into the cosmological horizon and consistent with the holographic principle. Second, the energy is introduced via a representative scale of the process, the Boltzmann constant $k_B$, which plays the role of a link between the geometrical representation of space $\mathbbm{R}^3$ with the macroscopic physical world.

\subsection{In the realm of electrochemistry: electromotive force and cell potential}

With the introduction of the concepts of anode, cathode, electrode, electrolyte, and ion, in 1835 by Michael Faraday, it was possible to describe phenomena, and give explanation, in the field of electrochemistry. 

The natural inclusion of entropy (or free energy) inside the fundamental equation of dynamics, open the door to treat a more general class of problems, particularly those related to the domain of electrochemistry. Taking Eq.~\ref{eq27} along a closed curve $\gamma$, with $\oint_{\gamma} (\mathbf{F}^{ext} \cdot d \mathbf{s})=q \cal{FE}$, and taking additionnally, $\Delta G=-q\cal{FE}$, according to the laws of electrolysis, and from Maxwell's equations $\Delta S=-(\frac{\partial \Delta G}{\partial T})_p$, then, from Eq.~\ref{eq26}, we obtain the Helmholtz formula for the electromotive force of a battery (or cell potential $\cal{E}$):
\begin{equation}
\mathcal{E}=U+T\frac{\partial \mathcal{E}}{\partial T}.
\end{equation}
The consistent integration of dynamics with thermodynamics permits addressing electrochemical processes via the modified Newton equation~\ref{eq27}.

\section{Conclusion}

The usual assumption of point particles when performing calculations in the frame of classical mechanics hinders the correct understanding of motion since in real world particles or bodies are extended objects, subject to deformation processes that alter their expected behavior. It was shown that from the gradient of rotational energy emerged the well-known pseudo-forces appearing in rotational moving frames.
The out of equilibrium technique proposed in this Letter may provide a new instrument for a more in-depth analysis for the dynamics of complex systems, in particular the emergence of complex behaviors.

\begin{acknowledgements}
The author gratefully acknowledge partial financial support by the International Space Science Institute (ISSI) in Bern, Switzerland, as visiting scientist, and express special thanks to Professor Roger-Maurice Bonnet, Dr. Edouard Mend\`{e}s Pereira, Prof. Rafael Rodrigo, Dr. Maurizio Falanga and Dr. Rui Ribeiro, for their friendship and support. We acknowledge kind permission to Springer Nature Group and Elsevier, to publish part of the material shown authored by myself and made public via Open Access.

\end{acknowledgements}



\end{document}